%% file: main.tex
\documentclass[a4, 11pt]{article}    
\usepackage{natbib}
\usepackage{setspace,float,rotating,multirow}
\usepackage{graphicx}
\usepackage{amssymb,amsmath}
\usepackage{tipa}
\usepackage{caption}
\usepackage{subcaption}
\usepackage{multirow}
\usepackage[utf8]{inputenc}
\usepackage{pdflscape}
\usepackage{afterpage}
\usepackage{adjustbox}
\usepackage{threeparttable}
\usepackage{array}
\usepackage{booktabs}
\usepackage{lscape}
\usepackage{tabularx} 
\usepackage{setspace}
\RequirePackage{color}
\RequirePackage[flushleft]{paralist}[2013/06/09]
\usepackage{amsthm}
\usepackage{epsfig}
\usepackage{pstricks-add}
\usepackage{threeparttable}

\RequirePackage{geometry}
\newlength{\qspace}
\newcounter{qnumber}
 {\vspace{\qspace}
  \begin{enumerate}[\bfseries 1\quad][10]%
    \setcounter{enumi}{\value{qnumber}}%
    \item%
 }
{
  \end{enumerate}
  \filbreak
  \stepcounter{qnumber}
 }

 {
  \begin{enumerate}[\bfseries (i)]%
    \setcounter{enumii}{#1}
    \addtocounter{enumii}{-1}
    \setlength{\itemsep}{5mm}
    \setlength{\parskip}{8pt}
 }
 {
  \end{enumerate}
 }

\title{Measurement Error Correction for Spatially Defined Environmental Exposures in Survival Analysis}
\author{Lin Ge, Ce Yang, David Zucker, Jiaxuan Li, Donna Spiegelman, Molin Wang}
\date{}

\theoremstyle{definition}

\begin{document}
\maketitle

\begin{abstract}

Environmental exposures are often defined using buffer zones around geocoded home addresses, but these static boundaries can miss dynamic daily activity patterns, leading to biased results. This paper presents a novel measurement error correction method for spatially defined environmental exposures within a survival analysis framework using the Cox proportional hazards model. The method corrects high-dimensional surrogate exposures from geocoded residential data at multiple buffer radii by applying principal component analysis for dimension reduction and leveraging external GPS-tracked validation datasets containing true exposure measurements. It also derives the asymptotic properties and variances of the proposed estimators. Extensive simulations are conducted to evaluate the performance of the proposed estimators, demonstrating its ability to improve accuracy in estimated exposure effects. An illustrative application assesses the impact of greenness exposure on depression incidence in the Nurses' Health Study (NHS). The results demonstrate that correcting for measurement error significantly enhances the accuracy of exposure estimates. This method offers a critical advancement for accurately assessing the health impacts of environmental exposures, outperforming traditional static buffer approaches.

\end{abstract}

\input{Sec1.tex}
\input{Sec2.tex}
\input{Sec3.tex}
\input{Sec4.tex}
\input{Sec5.tex}

\bibliography{refs}
\bibliographystyle{apalike}


\end{document}

%% file: Sec1.tex
\section{Introduction}









Neighborhood environment exposure is typically defined using buffer zones around geocoded home addresses, and it is often quantitatively assessed based on a specific radius or range of radii from the address. Common environmental exposures include air, water, soil, green space, and many modern environmental health studies \textbf{[cite MEEE papers]} focus on the impact of these environmental exposures on health outcomes. However, environmental interactions extend far beyond these static, spatially defined boundaries. Daily outdoor activities create dynamic exposure patterns, and failing to account for these variations can lead to biased findings. For example, while greenness (i.e., the area of green space) is linked to improved physical activity and mental health, such as reduced anxiety and depression \citep{berke2007protective,james2015review,banay2019greenness,wilt2023}, assessing these benefits solely based on home addresses can introduce measurement errors. In epidemiological studies, this issue is referred to as covariate measurement error. The true covariate $X$, known as the ``gold-standard" measurement, is often only approximated by the surrogate measurement $Z$, such as residential-based exposures in this context. Moreover, the surrogate environmental exposure may be high-dimensional due to the pre-defined buffer radii, compounding the challenges associated with covariate measurement error.


On the other hand, to study health outcomes such as disease incidence or mental health events, time-to-event data related to exposure variables is often available from scientific study designs. Cox proportional hazards models \citep{cox1972}, widely used in survival analysis, are effective for exploring the effect of exposures on the health outcomes and estimating hazard ratios. However, in the presence of measurement error in residential-based surrogate exposures, these conventional approaches risk diluting true effects, biasing estimates toward the null, and potentially overlooking critical environmental risk factors. 


For decades, it has been well established that measurement error and misclassification in exposures and outcomes lead to biased estimates of health effects, typically attenuating the results toward the null and increasing the likelihood of false-negative findings. \cite{prentice1982covariate} first addressed the covariate measurement error in Cox regression models by proposing a novel regression calibration approach, which assumes a linear Gaussian model for $X$ given $Z$, to correct for measurement error under specific assumptions. Subsequent studies have expanded on this, including \cite{Zucker}, who introduced a pseudo partial likelihood method for covariate errors, and \cite{liao2011survival}, who discussed the time-varying covariate error correction in survival analysis. However, few studies have explored the issue of high dimensional surrogates for a single exposure. \cite{Weller} developed a regression calibration approach for multiple surrogates in logistic regression, but there has been limited attention to this issue in Cox regression models. In our work, we focus on addressing this gap in the context of cox survival models. 


Our motivating example draws on two cohorts: main study cohort from the Nurses' Health Study (NHS) and an external validation cohort from the NHS III Mobile Health Substudy. The NHS is a nationwide prospective cohort of female nurses, enrolled in 1976 at ages 30 to 55, who have been followed biennially via mailed questionnaires. These questionnaires capture demographic characteristics, lifestyle factors, and health status, with response rates of around 90\% at each cycle  \citep{bao2016,hart2018}. Participants in this study reside in all U.S. states and the District of Columbia, and detailed residential address histories have been regularly updated and geocoded with latitude and longitude. The NHS III Mobile Health Substudy is a pilot study of NHS III, an on-going internet-based cohort that began in 2010, with similar aims to NHS \citep{fore2020,wilt2023}. The validation cohort tracks participants' mobility histories via GPS data, collected from 42 states across the U.S., with locations geocoded by latitude and longitude. The Landsat-derived Normalized Difference Vegetation Index (NDVI) was used to calculate residential-based greenness exposure for both NHS and NHS III Mobile Health Substudy cohorts, while GPS-tracked greeness exposure, considered the ``gold-standard", is only available for the NHS III Mobile Health Substudy. Both studies were approved by the Institutional Review Boards of Brigham and Women's Hospital, Boston, MA, and all participants provided consent for participation and the return of questionnaire surveys.


In this paper, we address the aforementioned challenges by proposing methods to correct one single exposure effect from high-dimensional surrogate measures using external validation data. The remainder of the paper is structured as follows: Section 2 introduces our proposed method and outlines the estimation and inference procedures for correcting measurement error. Section 3 investigates the finite sample properties of our proposed estimators through simulation studies based on main and validation study designs. Section 4 applies our method to a motivating environmental health example, examining the effect of greenness exposure on incidence of depression. Finally, Section 5 concludes with remarks on the implications of our findings.


%% file: Sec2.tex
\section{Methods}

\subsection{Notation and Preliminaries}

We consider a time-to-event survival data setting where the outcome for participant $i = 1, \ldots, N$ is denoted as $(T_i, D_i)$. For example, this could represent the time to onset of depression in an epidemiological study. The follow-up time $T_i$ is defined as the minimum of the failure time $T_i^0$, and the censoring time $C_i$, i.e., $T_i=\min(T_i^0, C_i)$, with the failure indicator $D_i$ defined as $D_i = I\{T_i=T_i^0\}$, where $I\{\cdot\}$ is the indicator function. Environmental exposure is measured by the Normalized Difference Vegetation Index (NDVI), which calculates the percentage of greenness within circular buffers around a participant's location at a given time. Let $X_i\in R$ represent the greenness exposure at a distance $r \in [0, R_x]$ from participant $i$. In this work, we do not address issues regarding the “uncertain geographic context” (i.e., the optimal choice of $R_x$), and instead assume a predetermined buffer, for example, $R_x$ = 30 meters.

This exposure data is available from Global Positioning System (GPS) measurements \citep{james2016spatial} for participants in the validation cohort. Let $\beta_1$ represent the effect of environmental exposure $X_i$. Note that the NDVI derived from GPS-tracked locations is considered the ``gold standard" measurement of greenness exposure, and we will refer to $X_i$ as GPS-based NDVI in the following context. 

Assuming that the exposure is related to the time-to-event outcome through a Cox proportional hazards (PH) model \citep{david1972regression}, the hazard function can be written as 
\begin{equation} \label{eq.1}
    \lambda_i (t; X) = \lambda_{i0} (t) \exp \left[\beta_1 X_i  + \pmb{\beta}_2^T \pmb{W}_{i} + \pmb{\beta}_3^T X_i\circ \pmb{W}_i \right]
\end{equation}
where $\lambda_0 (t)$ is the baseline hazard function, $\pmb{W}_{i}$ represents confounders (such as age, gender), and $\pmb{\beta}_2$ are their coefficients. To generalize the model, we include an interaction term, $X_i\circ \pmb{W}_i$, between the GPS-based NDVI $X_i$ and the confounders $\pmb{W}_i$ with coefficients $\pmb{\beta}_3$. 

The partial likelihood for this model is expressed as:
\begin{equation}\label{eq.2}
    L(\beta_1, \pmb{\beta}_2, \pmb{\beta}_3) = \prod_{i = 1}^N \left\{\frac{\exp[\beta_1 X_i  + \pmb{\beta}_2^T \pmb{W}_{i} + \pmb{\beta}_3^T X_i\circ \pmb{W}_i]}{\sum_{j=1}^N Y_j \exp[\beta_1 X_j  + \pmb{\beta}_2^T \pmb{W}_{j} + \pmb{\beta}_3^T X_j\circ \pmb{W}_j]}\right\}^{D_i}.
\end{equation}
where $Y_j = I(T_j \geq T_i)$. Solving the partial score equation $U(\beta_1, \pmb{\beta}_2, \pmb{\beta}_3) =  \partial \log L(\beta_1, \pmb{\beta}_2, \pmb{\beta}_3) / \partial [\beta_1, \pmb{\beta}_2^T, \pmb{\beta}_3^T]^T$ , and setting it equal to zero, yields point estimates for the parameters $\beta_1$, $\pmb{\beta}_2$, $\pmb{\beta}_3$.

\subsection{Approximation Based on Observed Surrogates in the Main Study}

However, GPS-based NDVI data is only available for participants in a validation study (VS) \citep{wilt2023}. In the main study (MS), only high-dimensional greenness exposure, measured as residential-based NDVI, is available. This exposure is calculated within several radial buffer zones around the residential address of each participant, potentially introducing measurement error due to participant mobility. As a result, the observed data in the main study is $\pmb{Z}_i = [Z_i(r_1), Z_i(r_2), \cdots Z_i(r_Z)]^T$, a surrogate for the true exposure $X_i$. Therefore, the MS data comprises: 
\begin{equation*}
    D_{MS} = \{\pmb{Z}_i, \pmb{W}_i, D_i, T_i: i = 1, \ldots, N\},
\end{equation*}
where the buffer radii $(r_1, r_2, \cdots, r_Z)$ are fixed values, arranged in increasing order (e.g., 90m, 150m, $\cdots$, 2100m in our example). These radii are typically larger than $R_x$ used in the GPS-based NDVI in the validation study. According to \cite{prentice1982covariate}, the induced hazard function can be expressed as a conditional expectation of the true exposure:
\begin{align}\label{eq.3}
    \lambda_i (t; Z) &= \lambda_{i0} (t) E \left\{ \exp \left[\beta_1 X_i  + \pmb{\beta}_2^T \pmb{W}_{i} + \pmb{\beta}_3^T X_i\circ \pmb{W}_i] \right] | \pmb{Z}_i, \pmb{W}_i, T_i \geq t \right\} \nonumber \\
    &\approx \lambda_{i0} (t) \exp[\pmb{\beta}_2^T \pmb{W}_i] E \left\{ \exp \left[\beta_1 X_i+ \pmb{\beta}_3^T X_i\circ \pmb{W}_i]  \right] | \pmb{Z}_i, \pmb{W}_i \right\},
\end{align}
where we assume: ($a$) the surrogate exposure has no predictive value given the true exposure, i.e., non-differential error with $\lambda_i (t, X_i) = \lambda_i (t, X_i, \pmb{Z}_i)$, and ($b$) the population at risk is representative of the disease-free population, implying a rare disease scenario where $\lambda_i (t, \pmb{Z}_i, T_i \geq t) = \lambda_i (t, \pmb{Z}_i)$. The partial likelihood is then approximated as follows.
\begin{equation}\label{eq.4}
    L(\beta_1, \pmb{\beta}_2, \pmb{\beta}_3) = \prod_{i = 1}^N \left\{\frac{ \exp[\pmb{\beta}_2^T \pmb{W}_i] E\{ \exp[\beta_1 X_i+\pmb{\beta}_3^T X_i\circ \pmb{W}_i  ] | \pmb{Z}_i, \pmb{W}_i \}}{\sum_{j=1}^N Y_j \exp[\pmb{\beta}_2^T \pmb{W}_j] E\{ \exp[\beta_1 X_j + \pmb{\beta}_3^T X_j\circ \pmb{W}_j  ] | \pmb{Z}_i, \pmb{W}_j\}}\right\}^{D_i},
\end{equation}
where the denominator accounts for the risk set just prior to time $T_i$. 

Since calculating the exact conditional expectation requires modelling the full distribution $p(X_i | \pmb{Z}_i, \pmb{W}_i)$, \cite{prentice1982covariate} and \cite{liao2011survival} discussed first- and second- order approximations, simplifying the likelihood by assuming moment conditions. In general, the cumulants $\kappa$ arising from the cumulant generating function approximation are the central moments, expressed as
\begin{equation*}
    \log E [\exp(\beta_1 X) | {Z}] = \beta_1 \kappa (1) +  \frac{\beta_1^{2} }{2!} \kappa (2) + \ldots = \beta_1 E(X | {Z}) + \frac{\beta_1^{2}}{2}  {\rm Var} (X |{Z}) + \ldots.
\end{equation*}

In a standard covariate measurement error context, the first-order approximation interchanges the exponential sign and the conditional expectation. The second-order approximation incorporates the conditional variance of the covariate errors, and is exact under the assumption of a multivariate normal distribution,
\begin{align*}
    X |{Z} \sim N(\mu ({Z}), \Sigma ({Z})), ~~
    E[ \exp (\beta_1 X ) | {Z} ] = \exp [\beta_1 \mu ({Z})  + \frac{\beta_1^{2} }{2!}\Sigma ({Z})].
\end{align*}

In our context, we define $\mu_i = E[X_i | \pmb{Z}_i, \pmb{W}_i]$, assuming that the second-order moment is negligible compared to the first-order moment. Therefore, the first-order approximation from the cumulant generating function gives
\begin{align}\label{eq.5}
    & E\{ \exp[\beta_1 X_i+\pmb{\beta}_3^T X_i\circ \pmb{W}_i  ] | \pmb{Z}_i, \pmb{W}_i \} = E\{ \exp[(\beta_1+\pmb{\beta}_3^T \pmb{W}_i) {X}_i ] | \pmb{Z}_i, \pmb{W}_i \}  
    \nonumber \\  & \approx \exp \left\{(\beta_1+\pmb{\beta}_3^T \pmb{W}_i)  E[{X}_i | \pmb{Z}_i, \pmb{W}_i]  \right\} ,
\end{align}
However, if the second or higher order moments are not negligible, asymptotic bias may arise in the estimation results. Next to evaluate $\mu_i = E[X_i | \pmb{Z}_i, \pmb{W}_i]$ in equation \eqref{eq.5}, we need a measurement error model for $\mu_i$, which we will discuss in next section, based on the validation study.

\subsection{Measurement Error Models}

Regardless of the assumptions made to simplify the partial likelihood, a measurement error model is required to specify the conditional expectation using an external or internal validation dataset,
\begin{align*}
    D_{EVS} &= \{X_{i}, \pmb{Z}_i, \pmb{W}_{i}: i = 1, \ldots, N_{V} \}; \\
    D_{IVS} &= \{X_{i}, \pmb{Z}_i, \pmb{W}_{i}, D_{i}, T_{i}: i = 1, \ldots, N_{V}\}.
\end{align*}

For the internal validation study (IVS), transportability naturally follows, as we observe the true exposure for a subset of participants from the main study. In the case of the external validation study (EVS), the transportability assumption is applied.

To better estimate $\mu_i = E[X_i | \pmb{Z}_i, \pmb{W}_i]$ using the surrogate measurements $\pmb{Z}_i$ from participants in the main study, we need to select a measurement error model that is appropriate for our specific context. A standard measurement error model of the following form is initially considered.
\begin{align}\label{eq.6}
    E[X_i | \pmb{Z}_i, \pmb{W}_i] = \alpha_{0} + \pmb{\alpha}_1^T \pmb{Z}_i   + \pmb{\alpha}_2^T \pmb{W}_{i} +\pmb{\alpha}_3^T \pmb{W}_{i}\circ \pmb{Z}_i. 
\end{align}
where $\pmb{\alpha}_1=[\alpha_{1, r_1},\alpha_{1, r_2},\cdots, \alpha_{1, r_Z}]^T$ and $\pmb{W}_{i}\circ \pmb{Z}_i$ represents the interaction terms between $\pmb{W}_{i}$ and $\pmb{Z}_i$. This model assumes an additive relationship between the GPS-based NDVI $X_i$ and the high-dimensional residential-based NDVI measurements $\pmb{Z}_i$, along with interaction effects.

Given the special characteristics of the real data in our motivating example, the multicollinearity assumption may be violated due to the high correlation among the residential-based surrogate NDVI measurements $\pmb{Z}_i$ across different buffer sizes $r$. To address this issue and eliminate redundant information, common techniques such as Principal Component Analysis (PCA) \citep{pearson1901, hotelling1933} and Restricted Cubic Splines (RCS) \citep{de1978} can be employed. 

In the PCA approach, we reduce the dimensionality of the multicollinear surrogate matrix $\pmb{Z}_i$. Let $\mathbf{L}$ be the loading matrix calculated from the residential-based NDVI measurements across all buffer sizes. The PCA transformation is written as $\pmb{Z}_i^{*} =\mathbf{L}\pmb{Z}_i$, and the measurement error model (MEM) approximated by selected principal components takes the following form, which is similar in formulation to the standard form in \eqref{eq.6}.
\begin{align}\label{eq.7}
   E[X_i | \pmb{Z}_i, \pmb{W}_i] \approx 
   \alpha_{0}^* + \pmb{\alpha}_1^{*T} \pmb{Z}_i^*   + \pmb{\alpha}_2^{*T} \pmb{W}_{i} +\pmb{\alpha}_3^{*T} \pmb{W}_{i}\circ \pmb{Z}_i^*.
\end{align} 

The RCS approach forms a similar approximation to \eqref{eq.7} with an adjustment using spline basis matrix. In the RCS approach, we generalize the coefficient vector $\pmb{\alpha}_1$ by expressing each component as $\alpha_{1,r}  = \sum_{k = 1}^{L_n} {\alpha}_{k}^* {\Gamma}_k(r) = \pmb{\Gamma}_r^T\pmb{\alpha_{1}}^*$, where $L_n$ is the number of splines in the RCS approach. The coefficient vector is written in matrix form as $\pmb{\alpha}_1 = [\pmb{\Gamma}_{r_1}^T\pmb{\alpha_{1}}^*,\pmb{\Gamma}_{r_2}^T\pmb{\alpha_{1}}^*,\cdots, \pmb{\Gamma}_{r_Z}^T\pmb{\alpha_{1}}^*]^T = \mathbf{\Gamma}^T\pmb{\alpha_{1}}^*$,
where $\mathbf{\Gamma}=[\pmb{\Gamma}_{r_1}, \cdots, \pmb{\Gamma}_{r_Z}]^T$ is the spline basis matrix and $\pmb{\alpha}_1^*\in R^{L_n}$ is the new coefficient vector replacing $\pmb{\alpha}_1$. The RCS transformation is thus $\pmb{Z}_i^{**} =\mathbf{\Gamma}\pmb{Z}_i$.


The next step is to estimate the coefficients needed to calculate the required moments of the conditional distribution, $\mu_i = E[X_i | \pmb{Z}_i, \pmb{W}_i]$, which will be used in the partial likelihood. This can be achieved using the ordinary least squares (OLS) to solve the following estimating equation (EE) for $\pmb{\alpha} = (\alpha_0, \pmb{\alpha}_1^T, \pmb{\alpha}_2^T, \pmb{\alpha}_3^T)$ or $\pmb{\alpha}^* =(\alpha_0^*, \pmb{\alpha}_1^{*T}, \pmb{\alpha}_2^{*T}, \pmb{\alpha}_3^{*T})$:
\begin{align}\label{eq.8}
    U^{O} (\pmb{\alpha}) = \sum_{i = 1}^{N_V} [1, \pmb{Z}_i^T, \pmb{W}_i^T, (\pmb{W}_i\circ \pmb{Z}_i)^T]^T (X_i - \mu_i) = 0.
\end{align}
Alternatively, a longitudinal design for the validation data could be analyzed using marginal analysis via a Generalized Estimation Equation (GEE) \citep{Zeger1986}, formulated as follows.
\begin{align}
    U^{V} (\pmb{\alpha}; \psi) = \sum_{i = 1}^{N_V} U_i^{V} (\pmb{\alpha}; \psi) = \sum_{i = 1}^{N_V} \frac{\partial {\mu_i}}{\partial \pmb{\alpha}}^T
    V_i^{-1}
    ({X_i} - {\mu_i}) = 0,
    \label{eq.9}
\end{align}
where $V_i$ is the working covariance of $X_i$, specified as either exchangeable or independence.

\subsection{Estimation under the MS/EVS Design}

Suppose we solve the estimating equations \eqref{eq.8} or \eqref{eq.9} and obtain the estimate $\hat{\pmb{\alpha}}$ (or $\hat{\pmb{\alpha}}^*$). These estimates can then be used to approximated the conditional mean required for the first-order approximation of the induced hazard function. In the main study and external validation study design (MS/EVS), following the measurement error models \eqref{eq.6} or \eqref{eq.7}, we estimate the conditional mean $\mu_i= E[X_i| \pmb{Z}_i, \pmb{W}_i; \pmb{\alpha}]$ as $\hat X_i = E[X_i | \pmb{Z}_i, \pmb{W}_i; \hat{\pmb{\alpha}}]$ for $i = 1, \ldots, N$. We will now proceed with discussing 
parameter estimation in the MS study.

For simplicity, let $\pmb{\beta} = (\beta_1, \pmb{\beta}_2^T, \pmb{\beta}_3^T)^T$, and define:
\begin{align*}
    \pmb{S}_E^{(k)} (\pmb{\beta}, \pmb{\alpha}, t) &= \frac{1}{N} \sum_{i=1}^{N} [\mu_i, \pmb{W}_i^T, (\mu_i\circ \pmb{W}_i)^T]^{T}_k Y_i\exp \left[\beta_1 \mu_i  + \pmb{\beta}_2^T \pmb{W}_i  + \pmb{\beta}_3^T \mu_i\circ \pmb{W}_i  \right],
\end{align*}
for $k = 0, 1, 2$, where $\pmb{u}_k=[\mu_i, \pmb{W}_i^T, (\mu_i\circ \pmb{W}_i)^T]^{T}_k$, with $\pmb{u}_{0} = 1$, $\pmb{u}_{1} = \pmb{u}$, and $\pmb{u}_{2} = \pmb{u} \pmb{u}^T$, and $\pmb{u}=[\mu_i, \pmb{W}_i^T, (\mu_i\circ \pmb{W}_i)^T]^{T}$. Substituting the parameter estimates $\hat{\pmb{\alpha}}$, we get
\begin{align*}
    \pmb{S}_E^{(k)} (\pmb{\beta}, \hat{\pmb{\alpha}}, t) &= \frac{1}{N} \sum_{i=1}^{N} [\hat{X}_i, \pmb{W}_i^T, (\hat{X}_i\circ \pmb{W}_i)^T]^{T}_k Y_i\exp \left[\beta_1 \hat{X}_i  + \pmb{\beta}_2^T \pmb{W}_i  + \pmb{\beta}_3^T \hat{X}_i\circ \pmb{W}_i  \right].
\end{align*}
The point estimates of the parameters in the MS study can then be obtained by maximizing the following partial likelihood:
\begin{align}\label{eq.10}
    L_E (\pmb{\beta}, \hat{\pmb{\alpha}}) &= \prod_{i = 1}^N \left\{\frac{ \exp[\pmb{\beta}_2^T \pmb{W}_i] \exp[ \beta_1\hat X_i + \pmb{\beta}_3^T \hat X_i\circ \pmb{W}_i] }{\sum_{j=1}^N Y_j \exp[\pmb{\beta}_2^T \pmb{W}_j ] \exp[\beta_1\hat X_j + \pmb{\beta}_3^T \hat X_j\circ \pmb{W}_j]}\right\}^{D_i}. 
\end{align}
The corresponding log-likelihood is
\begin{align}\label{eq.11}
    l_E (\pmb{\beta}, \hat{\pmb{\alpha}}) &= \sum_{i=1}^N D_i \left \{ \left[ \beta_1\hat X_i  + \pmb{\beta}_2^T \pmb{W}_i + \pmb{\beta}_3^T \hat X_i\circ \pmb{W}_i \right] - \log \sum_{j=1}^N Y_j \exp[\pmb{\beta}_2^T \pmb{W}_j] \exp \left[\beta_1\hat X_j + \pmb{\beta}_3^T \hat X_j\circ \pmb{W}_j \right ] \right \} \nonumber\\
    &= \sum_{i=1}^N D_i \left \{ \left[ \beta_1\hat X_i  + \pmb{\beta}_2^T \pmb{W}_i + \pmb{\beta}_3^T \hat X_i\circ \pmb{W}_i \right]- \log \pmb{S}_E^{(0)} (\pmb{\beta}, \hat{\pmb{\alpha}}, T_i) -  \log(1/N)\right \}.
\end{align}
The score equation for $\beta$ is then:
\begin{align}\label{eq.12}
    \pmb{U}_E (\pmb{\beta}, \hat{\pmb{\alpha}}) = \sum_{i = 1}^N \pmb{U}_{Ei} (\pmb{\beta}, \hat{\pmb{\alpha}})
    = \sum_{i = 1}^N D_i \left \{ [\hat{X}_i, \pmb{W}_i^T, (\hat{X}_i\circ \pmb{W}_i)^T]^{T} - \frac{\pmb{S}_E^{(1)} (\beta, \hat \alpha, T_i)}{\pmb{S}_E^{(0)} (\beta, \hat \alpha, T_i)} \right \} = 0,
\end{align}
which leads to the parameter estimates $\hat{\pmb{\beta}} = (\hat{\beta}_1, \hat{\pmb{\beta}}_2^T, \hat{\pmb{\beta}}_3^T)^T$.

\subsection{Approximate Consistency of the Resulting Estimator}

Generally, the point estimates obtained in the previous section may not be strictly consistent but are approximately consistent. Under certain conditions, $\hat{\pmb{\beta}}$ converges to a well-defined constant vector, $\pmb{\beta}^*$, which reasonably approximates the true value of $\pmb{\beta}$. To see this, we define $\pmb{s}^{(k)} (\pmb{\beta}, \pmb{\alpha}, t)$ as the expectation of $\pmb{S}^{(k)}(\pmb{\beta}, \pmb{\alpha}, t)$ with respect to the true model, as follows.
\begin{align*}
    \pmb{s}^{(k)} (\pmb{\beta}, \pmb{\alpha}, t) = E \left\{ \frac{1}{N} \sum_{i=1}^{N} Y_i [{X}_i, \pmb{W}_i^T, ({X}_i\circ \pmb{W}_i)^T]^{T}_k Y_i\exp \left[\beta_1 {X}_i  + \pmb{\beta}_2^T \pmb{W}_i  + \pmb{\beta}_3^T {X}_i\circ \pmb{W}_i  \right] | \pmb{Z}_i, \pmb{W}_i  \right\}
\end{align*}
for $k = 0, 1, 2$, and similarly, we define $\pmb{u}_{0} = 1$, $\pmb{u}_{1} = \pmb{u}$, and $\pmb{u}_{2} = \pmb{u} \pmb{u}^T$, where $\pmb{u} = [{X}_i, \pmb{W}_i^T, ({X}_i\circ \pmb{W}_i)^T]^{T}$. 

Under the following conditions \citep{struthers1986misspecified,lin1989robust},
\begin{itemize}
\item $\hat{\pmb{\alpha}} \to \pmb{\alpha}^*$ in probability for some $\pmb{\alpha}^*$;
\item There exist neighborhoods $A$ of $\pmb{\alpha}^*$ and $B$ of $\pmb{\beta}^*$ such that for any $T < \infty$, 
\begin{align*}
{\rm sup}_{t \in [0, T], \pmb{\beta} \in B, \pmb{\alpha} \in A} |\pmb{S}^{(k)} (\pmb{\beta}, \pmb{\alpha}, t) - \pmb{s}^{(k)} (\pmb{\beta}, \pmb{\alpha}, t) | \to 0
\end{align*}
in probability for $k = 0, 1, 2$. Moreover,
$\pmb{s}^{(0)} (\pmb{\beta}, \pmb{\alpha}, t)$ is bounded away from zero on $A \times B \times [0, T]$.
$\pmb{s}^{(k)} (\pmb{\beta}, \pmb{\alpha}, t)$ are continuous in $\pmb{\beta} \in B$ and $\pmb{\alpha} \in A$, and bounded on $A \times B \times [0, T]$ for $k = 0, 1, 2$;
\item For each $T < \infty$, $\int_0^{T} \pmb{s}^{(2)} (t) dt < \infty$,
\end{itemize}
the log-likelihood satisfies
\begin{align*}
    \frac{1}{N} l(\pmb{\beta}, \pmb{\alpha}) \to H(\pmb{\beta}, \pmb{\alpha}) = \int_0^{t^*} \pmb{\beta} \pmb{s}^{(1)} (t) - \pmb{s}^{(0)} (t) \log \pmb{s}^{(0)} (\pmb{\beta}, \pmb{\alpha}, t) dt
\end{align*}
in probability for $\pmb{\beta} \in B$ and $\pmb{\alpha} \in A$. The function $H(\pmb{\beta}, \pmb{\alpha})$ has first and second derivatives:
\begin{align*}
    \pmb{h}(\pmb{\beta}, \pmb{\alpha}) &= \int_0^{t^*} \pmb{s}^{(1)} (t) - \pmb{s}^{(0)} (t) \left [ \frac{\pmb{s}^{(1)} (\pmb{\beta}, \pmb{\alpha}, t)}{\pmb{s}^{(0)} (\pmb{\beta}, \pmb{\alpha}, t)} \right ] dt; \\
    \mathbf{I}(\pmb{\beta}, \pmb{\alpha}) &= -\int_0^{t^*} \pmb{s}^{(0)} (t) \left \{ \frac{\pmb{s}^{(2)} (\pmb{\beta}, \pmb{\alpha}, t)}{\pmb{s}^{(0)} (\pmb{\beta}, \pmb{\alpha}, t)} - \left [ \frac{\pmb{s}^{(1)} (\pmb{\beta}, \pmb{\alpha}, t)}{\pmb{s}^{(0)} (\pmb{\beta}, \pmb{\alpha}, t)} \right ]^2 \right \} dt .
\end{align*}
We further assume that $\mathbf{I}(\pmb{\beta}, \pmb{\alpha})$ is negative definite. Let $\pmb{\beta}^*$ be the solution to $\pmb{h}(\pmb{\beta}, \pmb{\alpha}^*) = 0$. The function $H(\pmb{\beta}, \pmb{\alpha}^*)$ is concave with a unique maximum at $\pmb{\beta} = \pmb{\beta}^*$.

On the other hand, $\hat{\pmb{\beta}}$ maximizes the log-likelihood $l(\pmb{\beta}, \hat{\pmb{\alpha}})$, which converges in probability to the log-likelihood $l(\pmb{\beta}, \pmb{\alpha}^*)$.
As a result, $\hat{\pmb{\beta}}$ converges in probability to the unique maximum $\pmb{\beta}^*$ of $H(\pmb{\beta}, \pmb{\alpha}^*)$, following the convex analysis arguments in \cite{andelsen1982cox}.

\subsection{Asymptotic Normality of the Resulting Estimator}

Using bivariate expansion from \eqref{eq.12},
\begin{align*}
    0 = \pmb{U}(\hat{\pmb{\beta}}, \hat{\pmb{\alpha}}) \approx \pmb{U}(\pmb{\beta}^*, \pmb{\alpha}^*) + \frac{\partial \pmb{U}(\pmb{\beta}^*, \pmb{\alpha}^*)}{\partial \pmb{\beta}} (\hat{\pmb{\beta}} - \pmb{\beta}^*) + \frac{\partial \pmb{U}(\pmb{\beta}^*, \pmb{\alpha}^*)}{\partial \pmb{\alpha}} (\hat{\pmb{\alpha}} - \pmb{\alpha}^*),
\end{align*}
we rearrange to obtain
\begin{align}\label{eq.13}
    \sqrt{N} (\hat{\pmb{\beta}} - \pmb{\beta}^*) &\approx \left [ - N^{-1} \frac{\partial \pmb{U}(\pmb{\beta}^*, \pmb{\alpha}^*)}{\partial \pmb{\beta}} \right ]^{-1} \frac{1}{\sqrt{N}} \left [ \pmb{U}(\pmb{\beta}^*, \pmb{\alpha}^*) + \frac{\partial \pmb{U}(\pmb{\beta}^*, \pmb{\alpha}^*)}{\partial \pmb{\alpha}} (\hat{\pmb{\alpha}} - \pmb{\alpha}^*) \right ].
\end{align}

Next we examine the terms on the right-hand side of \eqref{eq.13}.

Firstly, consider 
\begin{align}\label{eq.14}
    - \frac{1}{N} \frac{\partial \pmb{U}(\pmb{\beta}^*, \pmb{\alpha}^*)}{\partial \pmb{\beta}} = \frac{1}{N} \sum_{i=1}^N \int_0^{t^*} \left \{ \frac{\pmb{S}^{(2)} (\pmb{\beta}^*, \pmb{\alpha}^*, t)}{\pmb{S}^{(0)} (\pmb{\beta}^*, \pmb{\alpha}^*, t)} - \left [ \frac{\pmb{S}^{(1)} (\pmb{\beta}^*, \pmb{\alpha}^*, t)}{\pmb{S}^{(0)} (\pmb{\beta}^*, \pmb{\alpha}^*, t)} \right ]^2 \right \} N_i (dt) \to \mathbf{I}(\pmb{\beta}^*, \pmb{\alpha}^*)
\end{align}
in probability, as shown by \cite{andelsen1982cox}. Note that in the form of a stochastic integral, $N_i(dt) = I(T_i \leq t, D_i = 1)$ is a counting process for event occurrences at time $T_i$ for $i = 1, \ldots, N$, with
$N_i (dt) = 1$ if subject $i$ fails at time $t$, and 0 otherwise.

Secondly, using Theorem 2.1 from \cite{lin1989robust} and Theorem 3.2 from \cite{andelsen1982cox}, $\frac{1}{\sqrt{N}} \pmb{U}(\pmb{\beta}^*, \pmb{\alpha}^*)$ can be approximated by a sum of i.i.d. random vectors with terms of $o_p(1)$. As demonstrated in appendix of \cite{lin1989robust}, this is asymptotically equivalent to $\frac{1}{\sqrt{N}} \sum_{i=1}^N \pmb{G}_i (\pmb{\beta}^*, \pmb{\alpha}^*)$, where
\begin{align*}
    \pmb{G}_i (\pmb{\beta}, \pmb{\alpha}) &= \int_0^{t^*} \left \{ [{X}_i, \pmb{W}_i^T, ({X}_i\circ \pmb{W}_i)^T]^{T} - \frac{\pmb{s}^{(1)} (\pmb{\beta}, \pmb{\alpha}, t)}{\pmb{s}^{(0)} (\pmb{\beta}, \pmb{\alpha}, t)} \right \} N_i (dt) \\
    &- \int_0^{t^*} \frac{ Y_j \exp [ \beta_1 X_j + \pmb{\beta}_2^T \pmb{W}_j + \pmb{\beta}_3^T X_j\circ \pmb{W}_j]}{\pmb{s}^{(0)} (\pmb{\beta}, \pmb{\alpha}, t)} \left \{ [{X}_j, \pmb{W}_j^T, ({X}_j\circ \pmb{W}_j)^T]^{T} - \frac{\pmb{s}^{(1)} (\pmb{\beta}, \pmb{\alpha}, t)}{\pmb{s}^{(0)} (\pmb{\beta}, \pmb{\alpha}, t)} \right \} E\left[\frac{1}{N}\sum_{j=1}^N N_j (dt)\right]
\end{align*}
for $i = 1, \ldots, N$, which are i.i.d.
Thus, by the multivariate central limit theorem, 
\begin{align}    \label{eq.15}
    \frac{1}{\sqrt{N}} \pmb{U}(\pmb{\beta}^*, \pmb{\alpha}^*) \to N(0, E[\pmb{G}_i (\pmb{\beta}^*, \pmb{\alpha}^*)^2])
\end{align}
indicating convergence in distribution.

Thirdly, we account for the variability in estimating $\pmb{\alpha}$ using the VS dataset. A similar expansion as \eqref{eq.12} provides an approximation, 
\begin{align*}
    0 = \pmb{U}^V(\hat{\pmb{\alpha}}) \approx \pmb{U}^V (\pmb{\alpha}^*) + \frac{\partial \pmb{U}^V (\pmb{\alpha}^*)}{\partial \pmb{\alpha}} (\hat{\pmb{\alpha}} - \pmb{\alpha}^*) \approx \pmb{U}^V (\pmb{\alpha}^*) + E \left[ \frac{\partial \pmb{U}^V (\pmb{\alpha}^*)}{\partial \pmb{\alpha}} \right] (\hat{\pmb{\alpha}} - \pmb{\alpha}^*),
\end{align*}
and
\begin{align}\label{eq.16}
    \sqrt{N_V} (\hat{\pmb{\alpha}} - \pmb{\alpha}^*) &\approx \left \{ -\frac{1}{N_V} E \left [ \frac{\partial \pmb{U}^V (\pmb{\alpha}^*)}{\partial \pmb{\alpha}} \right ] \right\}^{-1} \frac{1}{\sqrt{N_V}} \pmb{U}^V (\pmb{\alpha}^*) \nonumber\\
    &\to N \left( 0, \left \{ \frac{1}{N_V} E \left [ \frac{\partial \pmb{U}^V (\pmb{\alpha}^*)}{\partial \pmb{\alpha}} \right ] \right\}^{-1} \frac{1}{N_V} E [\pmb{U}^V (\pmb{\alpha}^*)^2] \left \{ \frac{1}{N_V} E \left [ \frac{\partial \pmb{U}^V (\pmb{\alpha}^*)}{\partial \pmb{\alpha}} \right ] \right\}^{-1^T} \right).
\end{align}
Note that we also need to estimate $\psi$ when solving \eqref{eq.9}. Ad-hoc or method-of-moments approaches can be used. According to \citet{Liang1986}, the asymptotic normality for $\hat \alpha$ holds if $\hat \psi$ is $\sqrt{N_V}$-consistent given $\pmb{\alpha}$. One could substitute $\hat \psi$ to $\pmb{U}^V(\pmb{\alpha}^*)$ to estimate the asymptotic variance of $\hat \alpha$ as
\begin{align}\label{eq.17}
    \mathbf{V}(\pmb{\alpha}^*; \hat \psi) &=  \frac{1}{N_V} \left \{ \frac{1}{N_V} E \left [ \frac{\partial \pmb{U}^V (\pmb{\alpha}^*; \hat \psi)}{\partial \pmb{\alpha}} \right ] \right\}^{-1}  \frac{1}{N_V} E [\pmb{U}^V (\pmb{\alpha}^*; \hat \psi)^2] \left \{ \frac{1}{N_V} E \left [ \frac{\partial \pmb{U}^V (\pmb{\alpha}^*; \hat \psi)}{\partial \pmb{\alpha}} \right ] \right\}^{-1^T} .
\end{align}

In summary, combining the asymptotic properties derived from \eqref{eq.14}, \eqref{eq.15}, and \eqref{eq.17} into \eqref{eq.13}, we obtain the following result for the asymptotic distribution of $\hat{\pmb{\beta}}$:
\begin{align} \label{eq.18}
    &\sqrt{N} (\hat{\pmb{\beta}} - \pmb{\beta}^*) \to \nonumber\\
    & N \left(0, \left [\frac{1}{N} \frac{\partial \pmb{U}(\pmb{\beta}^*, \pmb{\alpha}^*)}{\partial \pmb{\beta}} \right ]^{-1} \left \{ E[\pmb{G}_i(\pmb{\beta}^*, \pmb{\alpha}^*)^2] + \frac{1}{N} \frac{\partial \pmb{U}(\pmb{\beta}^*, \pmb{\alpha}^*)}{\partial \pmb{\alpha}} \mathbf{V}(\pmb{\alpha}^*; \hat \psi)
     \left [\frac{\partial \pmb{U}(\pmb{\beta}^*, \pmb{\alpha}^*)}{\partial \pmb{\alpha}} \right ]^{T} \right \} \left [\frac{1}{N} \frac{\partial \pmb{U}(\pmb{\beta}^*, \pmb{\alpha}^*)}{\partial \pmb{\beta}} \right ]^{-1^T} \right).
\end{align}
since the covariance term 
\begin{align*}
    {\rm Cov} \left( \pmb{U}(\pmb{\beta}^*, \pmb{\alpha}^*), \frac{\partial \pmb{U}(\pmb{\beta}^*, \pmb{\alpha}^*)}{\partial \pmb{\alpha}} (\hat{\pmb{\alpha}} - \pmb{\alpha}^*) \right) &\approx E \left[ \pmb{U}(\pmb{\beta}^*, \pmb{\alpha}^*) \left\{ \frac{\partial \pmb{U}(\pmb{\beta}^*, \pmb{\alpha}^*)}{\partial \pmb{\alpha}} \left \{ -\frac{1}{N_V} E \left [ \frac{\partial \pmb{U}^V (\pmb{\alpha}^*)}{\partial \pmb{\alpha}} \right ] \right\}^{-1} \frac{1}{N_V} \pmb{U}^V (\pmb{\alpha}^*) \right\}^T \right] \\
    &- E[\pmb{U}(\pmb{\beta}^*, \pmb{\alpha}^*)] E \left\{ \frac{\partial \pmb{U}(\pmb{\beta}^*, \pmb{\alpha}^*)}{\partial \pmb{\alpha}} \left \{ -\frac{1}{N_V} E \left [ \frac{\partial \pmb{U}^V (\pmb{\alpha}^*)}{\partial \pmb{\alpha}} \right ] \right\}^{-1} \frac{1}{N_V} \pmb{U}^V (\pmb{\alpha}^*) \right\}^T \\
    &= E \left[\pmb{U}(\pmb{\beta}^*, \pmb{\alpha}^*) \left\{ \frac{\partial \pmb{U}(\pmb{\beta}^*, \pmb{\alpha}^*)}{\partial \pmb{\alpha}} \left \{ -\frac{1}{N_V} E \left [ \frac{\partial \pmb{U}^V (\pmb{\alpha}^*)}{\partial \pmb{\alpha}} \right ] \right\}^{-1} \frac{1}{N_V} \pmb{U}^V (\pmb{\alpha}^*) \right\}^T \right] \\
    &= 0.
\end{align*}
The second last line follows since $E[\pmb{U}(\pmb{\beta}^*, \pmb{\alpha}^*)] = 0$.
As for the last line, note that $\pmb{U}^V (\pmb{\alpha}^*)$ arises from the external validation dataset whereas $\pmb{U}(\pmb{\beta}^*, \pmb{\alpha}^*)$ arises from the main study dataset. 
So, $\pmb{U}^V (\pmb{\alpha}^*)$ is independent of both $\pmb{U}(\pmb{\beta}^*, \pmb{\alpha}^*)$ and $\partial \pmb{U}(\pmb{\beta}^*, \pmb{\alpha}^*) / \partial \pmb{\alpha}$. Since $E[\pmb{U}^V (\pmb{\alpha}^*)] = 0$, the covariance vanishes. The asymptotic index is $N \to \infty$, $N_V \to \infty$, and $N_V / N$ bounded by a finite number.
The asymptotic variance can be estimated as
\begin{align}\label{eq.19}
    \hat{{\rm Var}} (\hat{\pmb{\beta}}) = \frac{1}{N} \hat{\mathbf{I}}_{{\beta}}^{-1} \left[\hat{\mathbf{G}}_{\beta} + \frac{1}{N N_V}\hat{\pmb{U}}_{\alpha} \hat{\mathbf{V}}_{\alpha, \hat \psi} \hat{\pmb{U}}_{\alpha}^T \right] \hat{\mathbf{I}}_{\beta}^{-1^T},
\end{align}
where
{\small 
\begin{align*}
    \hat{\mathbf{I}}_{\beta}^{-1} &= \left [-\frac{1}{N} \sum_{i=1}^N \frac{\partial \pmb{U}_{Ei} (\pmb{\beta}, \hat{\pmb{\alpha}})}{\partial \pmb{\beta}} \right]_{\pmb{\beta} = \hat{\pmb{\beta}}}; \\
    \hat{\mathbf{G}}_{\beta} &= \left [\frac{1}{N} \sum_{i=1}^N \left \{ \pmb{U}_{Ei} (\pmb{\beta}, \hat{\pmb{\alpha}}) - \sum_{j=1}^N \frac{D_i Y_j e^{\beta_1 \hat{X}_j + \pmb{\beta}_2^T \pmb{W}_j + \pmb{\beta}_3^T \hat{X}_j\circ \pmb{W}_j }}{N \pmb{S}_E^{(0)} (\pmb{\beta}, \hat{\pmb{\alpha}}, T_j)} \left \{ [\hat{X}_i, \pmb{W}_i^T, \hat{X}_i\circ \pmb{W}_i^T]^{T} - \frac{\pmb{S}_E^{(1)} (\pmb{\beta}, \hat{\pmb{\alpha}}, T_j)}{\pmb{S}_E^{(0)} (\pmb{\beta}, \hat{\pmb{\alpha}}, T_j)} \right \} \right \}^2 \right]_{\pmb{\beta} = \hat{\pmb{\beta}}}; \\
    \hat{\pmb{U}}_{\alpha} &= \left [ \sum_{i=1}^N \frac{\partial \pmb{U}_{Ei} (\pmb{\beta}, {\pmb{\alpha}})}{\partial \pmb{\alpha}} \right ]_{\pmb{\beta} = \hat{\pmb{\beta}}, \pmb{\alpha} = \hat{\pmb{\alpha}}} = \left [ \frac{\partial}{\partial \pmb{\alpha}} \sum_{i=1}^N D_i \Biggl\{[\hat{X}_i, \pmb{W}_i^T, \hat{X}_i\circ \pmb{W}_i^T]^{T}- \frac{\pmb{S}_E^{(1)} (\pmb{\beta}, {\pmb{\alpha}}, T_i)}{\pmb{S}_E^{(0)} (\pmb{\beta}, {\pmb{\alpha}}, T_i)} \Biggl\}\right] _{\pmb{\beta} = \hat{\pmb{\beta}}, \pmb{\alpha} = \hat{\pmb{\alpha}}};\\
    \hat{\mathbf{V}}_{\alpha, \hat \psi} &= \frac{1}{N_V} \left [ \frac{1}{N_V} \sum_{i = 1}^{N_V} \frac{\partial \pmb{U}_i^V (\alpha; \hat \psi)}{\partial \alpha} \right ]^{-1}_{\pmb{\alpha} = \hat{\pmb{\alpha}}} \left [ \frac{1}{N_V} \sum_{i = 1}^{N_V} \pmb{U}_i^V (\pmb{\alpha}; \hat \psi)^2 \right ]_{\pmb{\alpha} = \hat{\pmb{\alpha}}} \left [ \frac{1}{N_V} \sum_{i = 1}^{N_V} \frac{\partial \pmb{U}_i^V (\pmb{\alpha}; \hat \psi)}{\partial \pmb{\alpha}} \right ]^{-1^T}_{\pmb{\alpha} = \hat{\pmb{\alpha}}}.
\end{align*}
}

%% file: Sec3.tex
\section{Simulation Studies}

In this section, we investigates the finite sample properties of our proposed estimators through simulation studies under the MS/EVS design, to simulate the real data analysis introduced in the next section. We consider a single true exposure covariate $X$ (e.g., GPS-based NDVI) with nine surrogate exposures $\pmb{Z}$ (e.g., residential-based NDVI) and a continuous error-free confounder ${W}$, mimicking the data structure of our motivating study. The true exposure $X$ is only available in the external validation study through repeated measurement data, while the surrogate exposures $\pmb{Z}$ are available in both the MS and EVS datasets. We compare the performance of the estimated coefficient, $\hat \beta_{1}$ of the true exposure $X$ in the outcome model of the main study, imputed using both the standard MEM approach as \eqref{eq.6} and Principal Component Analysis with three principal components as \eqref{eq.7}. In addition to the estimated coefficient of exposure $X$, we also evaluate the performance of the estimated coefficient $\hat{{\beta}}_{3}$ of the interaction term between $X$ and ${W}$. All simulation results are calculated based on 1000 simulated replicates. 

We examine two event rates in the simulation study: rare disease (event rate = 3.5\%) and common disease (event rate = 10\%), both inspired by the real dataset. The main study sample sizes are $n_1 = 10,000$ or $n_1 = 5,000$, while the validation study  sample sizes are $n_2 = 300$ or $n_2 = 150$. The validation study is generated with longitudinal observations across eight time points, under the standard MEM model \eqref{eq.6} to mimic the real validation data.

For the data simulation, we first generate the validation study data. The surrogate measures $\pmb{Z}$ are drawn from a multivariate normal distribution $MVN(\pmb{\mu}_Z, \mathbf{\Sigma}_{\pmb{\mu}_Z})$, and the confounder $W\sim N(1, 10)$ with a sample size of $8n_2$, where $\pmb{\mu}_Z$, $\mathbf{\Sigma}_{\pmb{\mu}_Z})$ are calculated from the empirical distribution of the residential-based NDVI measures in the validation study. The confounder distribution is based on adjusted Census tract population density divided by 1000. Two sets of coefficients were used for the standard MEM model to calculate $\hat X$: (1) $\alpha_0$ = 0.105, $\alpha_1$ = [0.184, 0.068, 0.290, -0.246, 0.311, -0.613, 0.381, 0.276, -0.103], $\alpha_2$ = -0.006, $\alpha_3$ = [-0.038, -0.080, -0.056, 0.215, -0.247, 0.309, -0.058, -0.121, 0.052] (mimicking the real data coefficients); (2) $\alpha_0 = 0.05$, $\alpha_1$ = [0.5,-0.5, 0.5, -0.5, 0.5, -0.5, 0.5, -0.5, 0.5], $\alpha_2$ = 0.1, $\alpha_3$ = [0.05, -0.05, 0.05, -0.05, 0.05, -0.05, 0.05,-0.05, 0.05]. In each case, the independence variance structure $\mathbf{\Sigma}_{X|Z,W} = \sigma_V^2 \mathbf{I}_{8n_2}$ is considered with three variances $\sigma_V^2 = 0.01$, 0.05 and 0.10. 

Next, for the main study data generation, the surrogate measures $\pmb{Z}$ and the confounder $W$ are generated similarly, with sample size $n_1$, and the MEM parameters remain the same. Two sets of coefficients considered in the main study's outcome model: (1) $\beta_1 = -0.284$, $\beta_2 = -0.049$, $\beta_3 = -0.047$ (mimicking the real data coefficients); (2) $\beta_1 = 1.0$, $\beta_2 = 0.1$, $\beta_3 = 0.1$. The event times are generated from a Weibull distribution with baseline hazard function $\lambda_0(t) = \theta\nu(\nu t)^{\theta-1}$, where $\theta = 10$ and $\nu=1$. The maximum follow-up time is generated from $t^*\sim unif(0, C_{max})$, where $C_{max}$ is chosen to achieve the desired event rate.


Table \ref{Table3} presents the simulation results to compare the performance of estimated exposure effects between the Standard Model (M1) and the PCA Model (M2) under the first setting that mimics the real data in our motivating example. The simulation examines bias, standard deviation (SD), standard error (SE), and 95\% confidence interval (CI) coverage across varying proportions of measurement error variance ($\sigma^2_V$) and sample sizes ($n_1, n_2$). Overall, the PCA model consistently demonstrates lower bias compared to the Standard model, particularly as variance increases in MEM. Especially, when $\sigma^2_V = 0.10$, the bias for the PCA model is substantially lower than for the Standard model across all sample size settings, suggesting that the PCA model is more robust to measurement error. However, the Standard model generally exhibits better efficiency across all settings as a trade-off of the cost of higher bias. Both models show comparable standard errors and maintain 95\% coverage in most scenarios. Notably, the coverage performance based on the derived asymptotic variance estimator is relatively unaffected by the increase in measurement error variance or changes in sample size.

\begin{table}
    \centering
    \begin{threeparttable}
    \caption{Simulation Results in Setting 1 Mimicking Real Data: Comparing the Performance of Estimated Exposure Effects for the Standard Model (M1) and PCA Model (M2) with Interaction Terms in the MEM Model $(\beta_{true}=-0.284)$}
    \label{Table3}
    \begin{tabular}{cccccccccccccc}
    \toprule
    \multirow{2}{*}{$p$} & \multirow{2}{*}{$n_1$} & \multirow{2}{*}{$n_2$} &   \multirow{2}{*}{$\sigma^2_V$} & \multicolumn{2}{c}{bias \%} & \multicolumn{2}{c}{SD} & \multicolumn{2}{c}{SE} 
      & \multicolumn{2}{c}{95\% CI} \\ 
    & &&&  M1& M2& M1& M2& M1& M2 & M1& M2  \\
    \hline
    \multirow{12}{*}{3.5\%} & \multirow{3}{*}{5000}  & \multirow{3}{*}{150} & 0.01 & 13.53 & 7.02  & 0.909 & 0.946 & 0.927 & 0.982 & 95.9 & 95.7 \\
                        &                        &                      & 0.05 & 15.99 & 5.10  & 0.825 & 0.940 & 0.949 & 1.032 & 97.7 & 96.5 \\
                        &                        &                      & 0.10 & 21.60 & 22.85 & 0.786 & 0.913 & 0.944 & 1.078 & 97.8 & 97.6 \\ \cline{2-12}
                        & \multirow{3}{*}{5000}  & \multirow{3}{*}{300} & 0.01 & 5.51  & 6.63  & 0.899 & 0.976 & 0.921 & 0.975 & 95.7 & 95.0 \\
                        &                        &                      & 0.05 & 19.99 & 7.11  & 0.893 & 0.973 & 0.943 & 1.006 & 96.6 & 95.7 \\
                        &                        &                      & 0.10 & 18.67 & 10.69 & 0.862 & 0.962 & 0.946 & 1.027 & 97.0 & 96.2 \\ \cline{2-12}
                        & \multirow{3}{*}{10000} & \multirow{3}{*}{150} & 0.01 & 3.28  & 4.34  & 0.624 & 0.691 & 0.653 & 0.694 & 96.5 & 95.5 \\
                        &                        &                      & 0.05 & 25.00 & 1.37  & 0.584 & 0.671 & 0.668 & 0.730 & 97.4 & 96.5 \\
                        &                        &                      & 0.10 & 36.23 & 23.33 & 0.547 & 0.654 & 0.662 & 0.767 & 97.8 & 98.1 \\ \cline{2-12}
                        & \multirow{3}{*}{10000} & \multirow{3}{*}{300} & 0.01 & 6.45  & 8.34  & 0.643 & 0.681 & 0.649 & 0.688 & 95.3 & 95.4 \\
                        &                        &                      & 0.05 & 11.25 & 0.61  & 0.605 & 0.695 & 0.663 & 0.708 & 97.1 & 94.9 \\
                        &                        &                      & 0.10 & 26.52 & 9.91  & 0.596 & 0.682 & 0.669 & 0.729 & 97.4 & 96.3 \\ \cline{1-12}
    \multirow{12}{*}{10\%} & \multirow{3}{*}{5000}  & \multirow{3}{*}{150} & 0.01 & 1.01  & 1.63  & 0.520 & 0.554 & 0.543 & 0.577 & 96.3 & 96.2 \\
                       &                        &                      & 0.05 & 17.70 & 10.69 & 0.491 & 0.568 & 0.559 & 0.603 & 97.6 & 96.5 \\
                       &                        &                      & 0.10 & 30.27 & 11.35 & 0.461 & 0.555 & 0.557 & 0.636 & 98.4 & 97.5 \\ \cline{2-12}
                       & \multirow{3}{*}{5000}  & \multirow{3}{*}{300} & 0.01 & 7.25  & 2.86  & 0.538 & 0.574 & 0.541 & 0.572 & 95.1 & 94.7 \\
                       &                        &                      & 0.05 & 11.69 & 1.23  & 0.520 & 0.557 & 0.551 & 0.587 & 96.2 & 96.6 \\
                       &                        &                      & 0.10 & 20.02 & 4.46  & 0.490 & 0.565 & 0.558 & 0.609 & 97.7 & 96.8 \\ \cline{2-12}
                       & \multirow{3}{*}{10000} & \multirow{3}{*}{150} & 0.01 & 1.95  & 1.00  & 0.373 & 0.404 & 0.384 & 0.408 & 96.0 & 94.5 \\
                       &                        &                      & 0.05 & 17.39 & 9.96  & 0.354 & 0.398 & 0.395 & 0.428 & 96.9 & 96.6 \\
                       &                        &                      & 0.10 & 31.18 & 7.38  & 0.330 & 0.378 & 0.392 & 0.452 & 97.2 & 97.7 \\ \cline{2-12}
                       & \multirow{3}{*}{10000} & \multirow{3}{*}{300} & 0.01 & 4.03  & 0.20  & 0.382 & 0.399 & 0.382 & 0.404 & 95.2 & 95.3 \\
                       &                        &                      & 0.05 & 10.53 & 1.93  & 0.359 & 0.399 & 0.391 & 0.415 & 96.7 & 95.9 \\
                       &                        &                      & 0.10 & 23.69 & 4.24  & 0.352 & 0.392 & 0.395 & 0.428 & 96.6 & 96.7 \\
    \bottomrule
    \end{tabular}
    \end{threeparttable}
\end{table}    

Meanwhile, Table \ref{Table4} demonstrates a similar pattern as observed in Table \ref{Table3}. The PCA model generally provides lower bias but loses some efficiency in setting 2, while the Standard model exhibits relatively better variability but at the cost of significantly higher bias. These trade-offs highlight the challenges of balancing bias and variance in exposure effect estimation, particularly when measurement error is non-negligible.

In addition to the simulation studies with the interaction terms included, the simulation comparisons for the Standard model (M3) and PCA model (M4) without interaction terms in the MEM model are presented in the Appendix (Table S1 and S2).

\begin{table}
    \centering
    \begin{threeparttable}
    \caption{Simulation Results in Setting 2: Comparing the Performance of Estimated Exposure Effects for the Standard Model (M1) and PCA Model (M2) with Interaction Terms in the MEM Model $(\beta_{true}=0.1)$}
    \label{Table4}
    \begin{tabular}{cccccccccccccc}
    \toprule
    \multirow{2}{*}{$p$} & \multirow{2}{*}{$n_1$} & \multirow{2}{*}{$n_2$} &   \multirow{2}{*}{$\sigma^2_V$} & \multicolumn{2}{c}{bias \%} & \multicolumn{2}{c}{SD} & \multicolumn{2}{c}{SE} 
      & \multicolumn{2}{c}{95\% CI} \\ 
    & &&&  M1& M2& M1& M2& M1& M2 & M1& M2  \\
    \hline
    \multirow{12}{*}{3.5\%} & \multirow{3}{*}{5000}  & \multirow{3}{*}{150} & 0.01 & 23.47  & 28.21 & 0.946 & 1.045 & 0.992 & 1.042 & 96.6 & 95.7 \\
    &  &   & 0.05 & 218.76 & 49.52 & 0.864 & 1.080 & 0.953 & 1.126 & 96.5 & 97.2 \\
    &  &   & 0.10 & 323.86 & 50.43 & 0.799 & 1.114 & 0.878 & 1.222 & 95.4 & 96.4 \\\cline{2-12}
    & \multirow{3}{*}{5000}  & \multirow{3}{*}{300} & 0.01 & 4.32   & 85.59 & 0.998 & 1.057 & 1.005 & 1.049 & 95.6 & 95.8 \\
    &  &   & 0.05 & 89.97  & 41.20 & 0.963 & 1.089 & 0.989 & 1.088 & 96.0 & 95.3 \\
    &  &  & 0.10 & 161.19 & 66.62 & 0.873 & 1.096 & 0.956 & 1.121 & 95.0 & 95.8 \\\cline{2-12}
    & \multirow{3}{*}{10000} & \multirow{3}{*}{150} & 0.01 & 72.92  & 16.46 & 0.664 & 0.734 & 0.693 & 0.732 & 96.2 & 95.5 \\
    &  &   & 0.05 & 246.87 & 19.16 & 0.613 & 0.818 & 0.652 & 0.788 & 93.7 & 95.5 \\
    &  &   & 0.10 & 359.55 & 39.36 & 0.566 & 0.821 & 0.612 & 0.868 & 91.8 & 96.0 \\\cline{2-12}
    & \multirow{3}{*}{10000} & \multirow{3}{*}{300} & 0.01 & 12.60  & 8.95  & 0.692 & 0.738 & 0.699 & 0.727 & 95.2 & 94.7 \\
    &  &   & 0.05 & 114.88 & 24.08 & 0.651 & 0.750 & 0.687 & 0.756 & 95.4 & 95.0 \\
    &  &   & 0.10 & 186.08 & 57.75 & 0.641 & 0.776 & 0.665 & 0.778 & 95.8 & 96.0 \\ \cline{1-12}
    \multirow{12}{*}{10\%} & \multirow{3}{*}{5000}  & \multirow{3}{*}{150} & 0.01 & 51.65  & 34.55  & 0.640 & 0.711 & 0.650 & 0.684 & 95.8 & 93.6 \\
    &  &   & 0.05 & 211.86 & 42.93  & 0.554 & 0.713 & 0.620 & 0.753 & 94.8 & 96.0 \\
    &  &   & 0.10 & 315.76 & 105.92 & 0.506 & 0.757 & 0.581 & 0.830 & 92.3 & 97.1 \\\cline{2-12}
    & \multirow{3}{*}{5000}  & \multirow{3}{*}{300} & 0.01 & 24.70  & 5.99   & 0.626 & 0.660 & 0.648 & 0.677 & 96.5 & 95.1 \\
    &   &  & 0.05 & 141.02 & 67.82  & 0.626 & 0.675 & 0.647 & 0.710 & 94.9 & 96.6 \\
    &   &  & 0.10 & 169.13 & 23.92  & 0.584 & 0.749 & 0.636 & 0.758 & 94.8 & 95.9 \\\cline{2-12}
    & \multirow{3}{*}{10000} & \multirow{3}{*}{150} & 0.01 & 73.18  & 8.17   & 0.444 & 0.469 & 0.454 & 0.480 & 96.1 & 95.8 \\
    &  &  & 0.05 & 219.56 & 38.77  & 0.407 & 0.514 & 0.441 & 0.527 & 92.4 & 96.6 \\
    &  &  & 0.10 & 311.63 & 48.65  & 0.383 & 0.524 & 0.418 & 0.579 & 87.3 & 96.5 \\\cline{2-12}
    & \multirow{3}{*}{10000} & \multirow{3}{*}{300} & 0.01 & 32.37  & 9.14   & 0.446 & 0.471 & 0.456 & 0.476 & 96.0 & 95.4 \\
    &    &   & 0.05 & 119.14 & 5.15   & 0.428 & 0.500 & 0.452 & 0.499 & 96.0 & 95.8 \\
    &   &   & 0.10 & 171.38 & 6.96   & 0.428 & 0.527 & 0.448 & 0.531 & 93.4 & 95.2 \\
    \bottomrule
    \end{tabular}
    \end{threeparttable}
\end{table} 

Analogue to Table \ref{table1} in real data analysis, we perform additional simulation study to examine the prediction metrics between the Standard model and PCA model for the simulated data in MEM model. The sample size for the validation study is set to $n_2=300$, which mimics for the real data scenario. Then a test data set is simulated with sample size $n_3=10,000$, which represents the prediction tasks for the main study data. Same metrics, MAE, MSE and QIC, are calculated for each simulated dataset to mimick the real data analysis in each run and the mean MAE, 25\% quantile MAE, 50\% (median) quantile MAE, 75\% quantile MAE, mean MSE and QIC are reported in Table \ref{Table5}.

Table \ref{Table5} shows the simulation results of each prediction metric of the Standard model and the PCA model. Overall, the PCA Model demonstrates better prediction performance, with lower errors and QIC values. The interaction terms show few effect on the prediction metrics for each model. Moreover, due to the high correlation among the residential-based NDVI measures at nine buffer radii, these results suggest that applying the PCA model can help improve prediction accuracy and model fit, even though the underlying generating model is based on the Standard model. 

\begin{table}
    \centering
    \begin{threeparttable}
    \caption{Simulation Results Comparing the Performance of Prediction Metrics between the Standard Model and PCA Model for Simulated Data$^{a}$}
    \label{Table5}
    \begin{tabular}{ccccccccccc}
    \toprule
     Model & Interaction & MAE & MAE (25\%) &  MAE (50\%)& MAE (75\%) & MSE& QIC  \\
    \hline
     Standard & No &  0.1848 & 0.1830 & 0.1847 & 0.1863 & 0.0537 & 56.98  \\  \hline
    PCA (3 PCs) & No &  0.1821 & 0.1808  & 0.1820 & 0.1832 & 0.0521 & 30.20  \\  \midrule
    Standard & Yes & 0.1848 & 0.1831 & 0.1847 & 0.1864 & 0.0537 & 49.88  \\  \hline
    PCA (3 PCs) & Yes &   0.1821 & 0.1808 & 0.1821 & 0.1832 & 0.0521 & 29.49    \\ 
    \bottomrule
    \end{tabular}
    \begin{tablenotes}
    \item[$^{a}$] \footnotesize True model in this simulation result is the Standard model with nine buffer radii and the interaction terms. The sample size for simulated validation data is $n_2=300$ and test set is $n_3=10000$, which mimics the prediction in main study data.
    \end{tablenotes}
    \end{threeparttable}
\end{table}

%% file: Sec4.tex
\section{Illustrative Example from Motivating Study}

As previously mentioned, our illustrative example is based on the MS/EVS design. The main study cohort consists of participants from the Nurse's Health Study (NHS), while the external validation study is drawn from NHS III Mobile Health Substudy. The objective of this analysis is to examine the effect of environmental greenness exposure on the incidence of depression among NHS participants. 

The main study cohort includes 41,945 women from the NHS who were alive, had no prior clinician-diagnosed depression and/or regular antidepressant use before 2000, and had at least one residential address geocoded within the contiguous U.S. between 2000 and 2016. Participants with a validated Mental Health Inventory-5 (MHI-5) score of $\leq$ 52 from the 1996 and/or 1998 NHS questionnaires were excluded due to probable depression \citep{annerstedt2012,cuijpers2009}. Additionally, those without available MHI-5 scores were excluded. The study ultimately comprised 449,139 person-years of follow-up, during which 5,454 incident cases of depression were identified. 

For the validation study, GPS-tracked data were collected from participants during four 7-day sampling periods across a year, using wearable devices and a customized smartphone application developed by Overlap Health, Inc., from 2018 to 2020. The final validation cohort includes longitudinal data from 335 participants, each recorded four times a year during 2018 and 2019. 

Greenness exposure, measured by the Normalized Difference Vegetation Index (NDVI), was derived from Landsat satellite data with a 30-meter resolution. This widely used environmental exposure marker in epidemiological studies \citep{james2015review,labib2020} quantifies the amount of photosynthesizing vegetation on land, with values ranging from -1 to 1. Higher values indicate greater coverage of photosynthesizing vegetation, while values below 0 indicate water and were set to 0 in the analysis. In this example, two types of NDVI calculations were used: residential-based NDVI and GPS-based NDVI. Residential-based NDVI was available for both MS and EVS cohorts and was calculated based on residential (mailing) addresses within nine circular radii buffer zones (90m, 150m, 270m, 510m, 750m, 990m, 1230m, 1500m, and 2100m). Smaller radii capture the immediate visual greenness around the residence, while larger radii reflect the broader distances people may walk. In contrast, GPS-based NDVI was measured only in the validation study, using a fixed 30-meter buffer radius around each participant based on GPS-tracked mobility points captured every 10 minutes. \citep{wilt2023}

\subsection{Measurement Error Model Selection}
In this illustrative example, our objective is to estimate the GPS-based NDVI greenness measurements using surrogate residential-based NDVI data at nine circular buffer radii within the validation dataset. By accurately predicting true greenness exposure for participants in the main study, we can assess the impact of greenness exposure on depression. 

We begin by selecting an appropriate measurement error model (MEM) based on the longitudinal data in the validation study. Given the repeated measurements for each participant, we use a GEE model for the MEM. Several potential MEMs are considered including: a standard MEM with 9 buffer radii, a standard MEM with a single buffer radius (using either 90m, 150m, 270m, or 510m), a Principal Component Analysis (PCA) approach with 2 or 3 principal components, a restricted cubic splines (RCS) approach with 3-7 default knots, and models incorporating all possible interaction terms. To identify the best-performing MEM, we employ a 5-fold cross-validation strategy, comparing models based on performance metrics such as mean absolute square error (MAE), mean squared error (MSE) and Quasi-likelihood under the Independence model Criterion (QIC), as proposed by \cite{pan2001}, which is analogous to the more commonly used AIC statistic. The results of the model selection process, comparing each potential MEM, are presented in Table \ref{table1}.

\begin{table}
    \centering
    \begin{threeparttable}
    \caption{Results for Model Selection to Compare Each Potential Measurement Error Model}
    \label{table1}
    \begin{tabular}{ccccccccccc}
    \toprule
    X$^{a}$ & Model$^{b}$ & Type & MAE & MAE (25\%) &  MAE (50\%)& MAE (75\%) & MSE& QIC  \\
    \hline
    \multirow{12}{*}{No}  & \multirow{5}{*}{Standard$^{c}$} & all included & 0.0881 & 0.0858 & 0.0893 & 0.0900 & 0.0130 & 154.08  \\ 
    & & only 90m & 0.0878 & 0.0873 & 0.0875 & 0.0892 & 0.0128 & 113.08  \\ 
    & & only 150m & 0.0867 & 0.0853 & 0.0864 & 0.0878 & 0.0126 & 110.13  \\ 
    & & only 270m & 0.0870 & 0.0818 & 0.0892 & 0.0903 & 0.0126 & 104.90  \\ 
    & & only 510m & 0.0878 & 0.0835 & 0.0901 & 0.0912 & 0.0129 & 108.10  \\ \cline{2-9}
    & \multirow{2}{*}{PCA} & 2 PCs & 0.0865 & 0.0844 & 0.0874 & 0.0889 & 0.0126 & 114.39  \\ 
    & & 3 PCs & 0.0869 & 0.0855 & 0.0872 & 0.0888 & 0.0127 & 121.26  \\ \cline{2-9}
    & \multirow{5}{*}{RCS} & 3 knots & 0.0867 & 0.0845 & 0.0874 & 0.0887 & 0.0126 & 122.63  \\ 
    & & 4 knots & 0.0876 & 0.0863 & 0.0876 & 0.0887 & 0.0128 & 129.63  \\ 
    & & 5 knots & 0.0879 & 0.0862 & 0.0894 & 0.0895 & 0.0129 & 132.42  \\ 
    & & 6 knots & 0.0879 & 0.0863 & 0.0895 & 0.0898 & 0.0129 & 137.24  \\ 
    & & 7 knots & 0.0879 & 0.0860 & 0.0892 & 0.0898 & 0.0130 & 142.84 \\ \midrule
    \multirow{15}{*}{Yes$^{d}$} & \multirow{5}{*}{Standard$^{c}$} & Race & 0.0889 & 0.0849 & 0.0895 & 0.0911 & 0.0133 & 169.00  \\ 
    & & popdnses & 0.0874 & 0.0836 & 0.0870 & 0.0912 & 0.0128 & 164.10  \\ 
    & & Age & 0.0885 & 0.0877 & 0.0890 & 0.0903 & 0.0130 & 199.50  \\ 
    & & nSES & 0.0892 & 0.0871 & 0.0895 & 0.0923 & 0.0133 & 211.00  \\ 
    & & married & 0.0890 & 0.0857 & 0.0901 & 0.0901 & 0.0132 & 201.50 \\ \cline{2-9}
    & & Race & 0.0868 & 0.0848 & 0.0886 & 0.0891 & 0.0127 & 121.50  \\ 
    & PCA & popdnses & 0.0871 & 0.0862 & 0.0870 & 0.0894 & 0.0128 & 123.20  \\ 
    & & Age & 0.0872 & 0.0845 & 0.0868 & 0.0890 & 0.0127 & 125.80  \\ 
    & (2 PCs) & nSES & 0.0868 & 0.0846 & 0.0873 & 0.0894 & 0.0126 & 130.00  \\ 
    & & married & 0.0865 & 0.0839 & 0.0868 & 0.0888 & 0.0125 & 126.10 \\ \cline{2-9}
    & & Race & 0.0874 & 0.0857 & 0.0883 & 0.0898 & 0.0128 & 128.20  \\ 
    & PCA & \textbf{popdnses} & \textbf{ 0.0864} & \textbf{0.0855} & \textbf{0.0859} & \textbf{0.0878} & \textbf{0.0125} & \textbf{118.10}  \\ 
    & & Age & 0.0877 & 0.0864 & 0.0868 & 0.0886 & 0.0129 & 138.20  \\ 
    & (3 PCs) & nSES & 0.0873 & 0.0856 & 0.0872 & 0.0892 & 0.0128 & 146.00  \\
    & & married & 0.0869 & 0.0858 & 0.0863 & 0.0885 & 0.0127 & 137.70 \\ 
    \bottomrule
    \end{tabular}
    \begin{tablenotes}
    \item[$^{a}$] \footnotesize X indicates whether interaction terms are included in the MEM model.
    \item[$^{b}$] \footnotesize The standard MEM model is introduced in Eq.\eqref{eq.6}, while Principal component analysis (PCA) and restricted cubic splines (RCS) are introduced in Eq.\eqref{eq.7}.
    \item[$^{c}$] \footnotesize The standard MEM model without interaction terms considers five different types: adjusting for all nine buffer radii (all included), or adjusting for only one buffer radius at 90m, 150m, 270m, 510m respectively. Meanwhile, the standard MEM model with interaction terms evaluates one type (with all nine buffer radii included) and includes all interaction terms between each NDVI measure and one covariate.
    \item[$^{d}$] \footnotesize The interaction terms are evaluated for each covariate, including Race, Census tract population density (popdnses), Age, race, Census tract (neighborhood) socioeconomic status (nSES), and marital status (married).
    \end{tablenotes}
    \end{threeparttable}
\end{table}

In Table \ref{table1}, the potential models are categorized based on whether interaction terms are included. Initially, we applied the GEE model without interaction terms and found that the PCA approach outperformed others in terms of both MAE and MSE. As a result, we focused on the PCA approach when examining interaction terms, while also benchmarking it against the standard MEM model with nine buffer radii, a commonly used approach in practice. Ultimately, the optimal MEM model selected was the PCA approach with an interaction term for Census tract population density (popdnses), as it demonstrated the lowest MAE and MSE across all models. This MEM model will be used for the main study data analysis.

\subsection{Results for Real Data Analysis}
After selecting the optimal MEM model from the validation data, we imputed the predicted GPS-based NDVI measurements for each participant in the main study dataset. These imputed values were then used in a Cox proportional hazards model to assess the association between greenness exposure and depression outcomes in the main study. The results of this analysis are presented in table \ref{table2}.

\begin{threeparttable}
\centering
\caption{Hazard ratio (HR) and 95\% confidence intervals (CI) per 0.1 increase in mobility-based NDVI with depression incidence and effect modified by the census tract population density in the Nurses' Health Study$^{a}$}
\label{table2}
\begin{tabular}{lllcccccc}
\toprule
\multirow{2}{*}{Setting$^{b}$} & \multirow{2}{*}{Models$^{c}$} &    \multicolumn{3}{c}{Census Tract Population Density$^{d}$}  \\
 && at 0 people/sq mi & at 604 people/sq mi&  at 1,094  people/sq mi \\
\hline
\multirow{2}{*}{MEM w/o int} & Standard Model  & 0.965 [0.916, 1.016] & 0.956 [0.907, 1.008] & 0.949 [0.899, 1.002] \\
& PCA Model &  0.962 [0.913, 1.015]  & 0.954 [0.904, 1.007] &  0.947 [0.896, 1.001] \\
\multirow{2}{*}{MEM w/ int} & Standard Model &  0.963 [0.919, 1.009] & 0.962 [0.918, 1.007] & 0.960 [0.917, 1.006] \\
& PCA Model &  0.954 [0.905, 1.005]  & \textbf{ 0.947 [0.898, 1.000] } & \textbf{0.942 [0.892, 0.996]}  \\
\bottomrule
\end{tabular}
\begin{tablenotes}
    \item[$^{a}$] \footnotesize Hazard ratios are adjusted for age, race, marital status, education attainment, husband's education attainment, Census tract population density and its interaction with NDVI, neighborhood socioeconomic status (nSES), and fine particulate matter air pollution (PM2.5).
    \item[$^{b}$] Two models are compared: the MEM without the interaction term for Census tract population density (``MEM w/o int") and the MEM with the interaction term for the Census tract population density (``MEM w/ int").
    \item[$^{c}$] The standard model is calculated using \eqref{eq.6}, while the PCA model is based on \eqref{eq.7} with first three principal components.
    \item[$^{d}$] Census tract population density in the NHS cohort is treated as a continuous variable, with a median of 604 people per square mile and an average of 1,094 people per square mile. Population densities over 1,000 people per square mile typically indicate urban areas. \citep{coburn2007}
\end{tablenotes}
\end{threeparttable}

In Table \ref{table2}, the two MEM model settings are compared, with the MEM model that includes interaction terms being preferred based on the model selection in Table \ref{table1}. The hazard ratio for depression incidence per 0.1 increase in NDVI greenness exposure is 0.947 [0.898, 1.000] at the median Census tract population density of the NHS cohort (604 people per square mile). This effect is more pronounced at the mean population density (1,096 people per square mile), yielding a hazard ratio of 0.942 [0.892, 0.996]. Areas with a population density exceeding 1,000 people per square mile are classified as urban \citep{coburn2007}, suggesting that greenness exposure is associated with a more significant reduction in depression incidence in urban area compared to rural ones.

%% file: Sec5.tex
\section{Discussion}

In this paper, we developed a novel method to correct measurement errors in spatially defined environmental exposures, focusing specifically on the effects of greenness exposure on the incidence of depression within a survival analysis framework. The method leverages high-dimensional surrogate measures, such as residential-based greenness exposure, and corrects them using external validation data through measurement error models. This approach addresses a key challenge in environmental health studies that the bias introduced by assuming residential greenness exposure reflects individual physical patterns. By incorporating GPS-based greenness exposure from a smaller validation dataset, we correct for these measurement errors.

The proposed method was applied to real data from the Nurses' Health Study (NHS) study, investigating the relationship between greenness exposure and depression incidence in a long-term cohort. The results underscored that failing to account for measurement error can substantially bias the estimated risk toward the null, masking the protective effects of greenness on depression incidence. By implementing our correction method, we found a more pronounced association between increased greenness and a reduced risk of depression, particularly in urban area with higher population densities. These findings highlight the importance of measurement error correction in accurately assessing environmental health impacts.

Simulation studies further demonstrated the robustness of our method across varying sample sizes and degrees of measurement error. Notably, when using principal component analysis (PCA) for dimensionality reduction in surrogate exposures, our method consistently produced lower bias compared to standard approaches. However, the results also indicated a trade-off between bias and efficiency, with the standard model exhibiting greater efficiency but at the cost of increased bias.

In conclusion, this study provides a robust framework for addressing measurement error in spatially defined environmental exposures. Correcting these errors allows for more accurate estimates of exposure effects on health outcomes, which is critical for informing public health policies and interventions aimed at mitigating the impacts of environmental risk factors. Future research could extend this method to other environmental exposures and explore the effects of more complex interaction terms.